\documentclass[twocolumn,showpacs,preprintnumbers,amsmath,amssymb]{revtex4-1} \usepackage{dcolumn}

\usepackage{graphicx} \usepackage{bm}

\newcommand{\hs}{\hat{s}}

\newcommand{\hAe}{\hat{A}^{e}}
\newcommand{\hAo}{\hat{A}^{o}}
\newcommand{\hchi}{\hat{\chi}}

\newcommand{\hH}{\hat{H}}

\newcommand{\hI}{\hat{I}}
\newcommand{\hK}{\hat{K}}

\newcommand{\hP}{\hat{P}}
\newcommand{\hPe}{\hat{P}^{e}}
\newcommand{\hPo}{\hat{P}^{o}}
\newcommand{\hPi}{\hat{\Pi}}
\newcommand{\hS}{\hat{S}}
\newcommand{\hSig}{\hat{\Sigma}}
\newcommand{\hT}{\hat{T}}

\newcommand{\chP}{\hat{\cal P}}
\newcommand{\cSe}{{\cal S}^e}
\newcommand{\cSo}{{\cal S}^o}
\newcommand{\cSeta}{{\cal S}^{\eta}}

\begin{document}

\title{The Pauli Exclusion Operator: example of Hooke's atom}
\date{\today}
\author{Tomasz M. Rusin } \email{email: tmr@vp.pl} \author{Wlodek Zawadzki}
\affiliation{Institute of Physics, Polish Academy of Sciences,
 Al. Lotnik\'ow 32/46, 02-688 Warsaw, Poland}

\begin{abstract}
The Pauli Exclusion Operator (PEO) which ensures
proper symmetry of the eigenstates of multi-electron systems with
respect to exchange of each pair of electrons is introduced.
Once PEO is added to the Hamiltonian, no additional constraints on
multi-electron wave function due to the Pauli exclusion principle are needed.
For two-electron states in two dimensions ($2D$)
the PEO can be expressed in a closed form in terms of momentum operators, while
in the position representation PEO is a non-local operator.
Generalizations of PEO for multi-electron systems is introduced.
Several approximations to PEO are discussed.
Examples of analytical and numerical calculations of PEO
are given for isotropic and anisotropic Hooke's atom in~$2D$.
Application of approximate and kernel forms of PEO for calculations of
energies and states in~$2D$ Hooke's atom are analyzed.
Relation of PEO to standard variational calculations with the use of Slater
determinant is discussed.
\end{abstract}

\maketitle

\section{Introduction \label{Sec_Intro}}

Two-electron systems, e.g. the helium atom, were analyzed from the early years of
quantum mechanics~\cite{Kellner1927,Hylleraas1928,Hylleraas1929}.
Since the exact solutions of such situations are not known one
usually calculates the energies of low states and the corresponding wave functions
using the variational method. To be consisted with the Pauli exclusion
principle~\cite{Pauli1925,Pauli1940} one
first selects the spin state of the electron pair, that is either a singlet or a triplet,
and then assumes the trial functions of two
electrons to be either symmetric or antisymmetric
with respect to exchange of the two particles. This approach was successfully applied to
the ground energy of the helium atom as well as to its excited
states~\cite{BetheBook,Tanner2000}.

The Pauli exclusion principle can be introduced
to the variational calculations by choosing the trial function
of required symmetry with respect to exchange of the electrons.
This approach may not be used in a numerical integration of the
Schrodinger equation of the two electron systems since
this equation does not include terms which can be related
to the Pauli exclusion principle. Then, if one integrates
this equation for two electrons or for two non-fermions
having the same charges and masses as the electrons,
then in both cases one obtains the same
energies and states.

However, for the two-electron case some calculated states do
not fulfill the Pauli exclusion principle and such states
have to be eliminated as nonphysical ones. As an example,
wave functions symmetric with respect to exchange of electrons
are allowed for the singlet, but have to be eliminated for the triplet.

One can then state that, beyond the external potential and the Coulomb
repulsion, there exists an additional spin-dependent field
acting on both electrons
which eliminates some states from the spectrum of the Hamiltonian~$\hH$.
The presence of this field can be included in the model by introducing
a spin-depended operator~$\hP$ responsible for the existence of
the Pauli exclusion principle.
The final effect of the operators~$\hH$ and~$\hP$ acting on the
eigenstate~$|\Psi(1,2)\rangle$ of~$\hH$ is that
the states of the proper electron exchange symmetry
are not altered but those of the improper symmetry vanish.
Then, by solving numerically the Schrodinger equation
with the operator~$\left(\hH - \hP\right)$ instead of~$\hH$
\begin{equation}
 \left( \hH - \hP \right) |\Psi(1,2)\rangle = E |\Psi(1,2)\rangle,
\end{equation}
one automatically obtains states
fulfilling the Pauli exclusion principle, and
no additional constrains on multi-electron wave function
due to the Pauli exclusion principle are needed.
The main purpose of this work is to analyze the operator~$\hP$,
[called further the Pauli Exclusion Operator (PEO)],
in several two-electron systems. We show that in
these cases it is possible to obtain PEO in a closed form.
We also discuss generalization of PEO for multi-electron case and
propose several approximations of this operator.
Note that PEO exists in the literature in a different meaning and
it was used to calculate nuclear
matter~\cite{Cheon1989,Schiller1999,Suzuki2000}, see Discussion.

It is impossible to obtain PEO
for the helium atom because of two reasons. First, the Schrodinger equation
of the latter does not separate into a sum of two one-electron equations,
so one has to solve numerically the eigenequation
in the six-dimensional space. Second, in the presence of the attractive
Coulomb potential of helium nucleus there exist both localized and
delocalized electron states, and the latter are difficult
to be treated numerically.

There exists a model in which one avoids the above problems.
This system, called the Hooke's atom, consists of two electrons in the
field of~$N$-dimensional harmonic
oscillator~\cite{Kais1989,Taut1993,Taut1994,ONeil2003,HookesWiki}.
In this model the Schrodinger
equation separates into two equations of the center-of-mass
and relative motion of electrons. For potentials with a radial
symmetry one obtains two one-dimensional equations
which are much easier to solve numerically.
For sufficiently strong harmonic potential the spectrum of the Hooke's atom
consists of the localized states alone. For these reasons we analyze here
PEO in Hooke's atom model and then generalize obtained results for
multi-electron case.

The work is organized as follows. In Section~\ref{Sec_Theory}
we introduce the Pauli Exclusion Operator for~$2D$ two-electron systems.
In Section~\ref{Sec_MultiE} we generalize PEO for multi-electron systems
and propose several approximations of PEO.
In Section~\ref{Sec_Examples} we show examples of PEO in two~$2D$ Hooke's
atoms and calculate them analytically and numerically. In the same section
we show examples of approximate formulas for PEO.
In Section~\ref{Sec_Discussion} we discuss the obtained results,
while in the appendices we describe a numerical
method of obtaining low and high energy
states of the Hooke's atom and provide auxiliary formulas.
The work is concluded by the Summary.

\section{Two-electron systems~\label{Sec_Theory}}

In the atomic units the Hamiltonian of two interacting electrons
in the presence of an external potential~$U({\bm r})$ reads
\begin{equation} \label{ThHU}
 \hH = -\frac{1}{2} {\bm \nabla}_1^2 -\frac{1}{2} {\bm \nabla}_2^2 +
 \frac{e^2}{|{\bm r}_1 - {\bm r}_2|} + U({\bm r}_1) + U({\bm r}_2).
\end{equation}
We consider a~$2D$ case.
The description given in Eq.~(\ref{ThHU}) is not complete
because the solutions have to be limited to those fulfilling
the Pauli exclusion principle. The two-electron wave
function~$\Psi({\bm r}_1, {\bm r}_2)$, being the eigenstate of~$\hH$
should be either symmetric (for the singlet state)
or antisymmetric (for triplet states) with respect to
exchange~${\bm r}_1 \Leftrightarrow {\bm r}_2$.
We introduce the center-of-mass~${\bm R} = ({\bm r}_1 + {\bm r}_2)/2$ and the
relative motion~$\bm r = {\bm r}_1 - {\bm r}_2$.
In the new coordinates the exchange
of electrons does not affect~${\bm R}$ but changes sign of~${\bm r}$,
i.e.~${\bm r} \rightarrow -{\bm r}$. Then there is
\begin{equation} \label{ThRr}
 \Psi({\bm R},-{\bm r}) = \left\{\begin{array}{rr} \Psi({\bm R},{\bm r}) & \textrm{for singlet},
 \\ -\Psi({\bm R},{\bm r}) & \textrm{for triplets}. \end{array} \right.
\end{equation}
In the circular coordinates~${\bm r} = (r,\phi)$ the change~${\bm r} \rightarrow -{\bm r}$
corresponds to the transformation:~$(r,\phi) \rightarrow (r,\phi+\pi)$.
We introduce symmetric (even in~${\bm r}$)
and anti-symmetric (odd in~${\bm r}$) parts of~$\Psi({\bm R}, r,\phi)$
\begin{eqnarray}
 \label{ThPsie}
 \Psi({\bm R}, r,\phi)^e = \frac{1}{2} \left[ \Psi({\bm R}, r,\phi) + \Psi({\bm R}, r,\phi + \pi) \right], \\
 \label{ThPsio}
 \Psi({\bm R}, r,\phi)^o = \frac{1}{2} \left[ \Psi({\bm R}, r,\phi) - \Psi({\bm R}, r,\phi + \pi) \right].
\end{eqnarray}
Because of the existence of the Pauli exclusion principle one obtains two separate
eigenproblems for~$\Psi({\bm R}, r,\phi)^{\eta}$ (with~$\eta \in\{e,o\}$)
\begin{equation} \label{ThPsieta}
 \hH \Psi({\bm R}, r,\phi)^{\eta} = E^{\eta} \Psi({\bm R}, r,\phi)^{\eta},
\end{equation}
instead of the single problem for~$\Psi({\bm R}, r,\phi)$.
We can introduce the spin-dependent operator~$\hP$,
which we call the Pauli Exclusion Operator (PEO),
which for a given combination of electron spins removes
even or odd states from the spectrum of~$\hH$.
We define~$\hP$ as, see Eq.~(\ref{ThPsieta})
\begin{equation} \label{ThPs}
\left(\hH - \hP\right) \Psi({\bm R}, r,\phi) = \hH \Psi({\bm R}, r,\phi)^e,
\end{equation}
for a symmetric function of spins~$\hs_1,\hs_2$, and
\begin{equation} \label{ThPa}
 \left(\hH - \hP\right) \Psi({\bm R}, r,\phi) =\hH \Psi({\bm R}, r,\phi)^o,
\end{equation}
for antisymmetric function of~$\hs_1,\hs_2$.
In Eqs.~(\ref{ThPs}) and~(\ref{ThPa}) the operator~$\left(\hH - \hP\right)$
acts on~$\Psi({\bm R}, r,\phi)$, while the operator~$\hH$
in Eq.~(\ref{ThPsieta}) acts on~$\Psi({\bm R}, r,\phi)^{\eta}$.
In their spectrums the operators~$\hP$ and~$\left(\hH - \hP\right)$ contain
states having opposite symmetry with respect to a
change~${\bm r} \rightarrow -{\bm r}$, and sets of states belonging to
both operators are disjointed.
A closed form of~$\hP$ for multi-electron systems is unknown, but
for two-electron Hamiltonians in~$2D$ we can express~$\hP$ in
terms of differential operators and as a nonlocal operator
in the position representation.

To find the spectrum of~$\hP$ we introduce two auxiliary operators~$\hPe$ and~$\hPo$.
Let~$\hPe$ equals~$\hP$ in Eq.~(\ref{ThPs}) and~$\hPo$ in Eq.~(\ref{ThPa}).
Let~$|{\rm n}\rangle$ and~$E_{\rm n}$ be the states and energies of~$\hH$, respectively.
Then~$\hH = \sum_{\rm n} E_{\rm n} |{\rm n}\rangle\langle {\rm n}|$, and
\begin{eqnarray}
 \label{ThPe}
 \hPe &=& \sum_{{\rm n}\ \textrm{even}} E_{\rm n} |{\rm n}\rangle\langle {\rm n}|, \\
 \label{ThPo}
 \hPo &=& \sum_{{\rm n}\ \textrm{odd}} E_{\rm n} |{\rm n}\rangle\langle {\rm n}|,
\end{eqnarray}
where 'even' and 'odd' means that in the summations we restrict ourselves to
states being even or odd functions of~${\bm r}$, respectively. The above form of
operators~$\hPe$ and~$\hPo$ is useful if one knows all energies and
states of~$\hH$. Examples of such calculations are presented in the next section.
The operators~$\hPe$ and~$\hPo$ are on the same order as~$\hH$
and they may not be treated as perturbations to~$\hH$.
Operator~$\hP$ depends on the Hamiltonian of the system.

On the left sides of Eqs.~(\ref{ThPs}) and~(\ref{ThPa}) there is
the function~$\Psi({\bm R}, r,\phi)$ while on the right sides
there are functions~$\Psi({\bm R}, r,\phi)^e$ or~$\Psi({\bm R}, r,\phi)^o$.
To find a more symmetric form of these equations let us
insert~$\hPe$ in Eq.~(\ref{ThPe}) into Eq.~(\ref{ThPs}). Then one has
\begin{eqnarray} \label{ThhHhPeX}
 \left(\hH - \hPe \right)|\Psi\rangle &=&
 \sum_{\rm n} E_{\rm n} |\rm n\rangle \langle {\rm n}|\Psi \rangle -
 \sum_{{\rm n}\ \textrm{even}} E_{\rm n}|{\rm n}\rangle\langle {\rm n}|\Psi \rangle \nonumber \\
 &=&\sum_{{\rm n}\ \textrm{odd}} E_{\rm n}|{\rm n}\rangle\langle {\rm n}|\Psi \rangle.
\end{eqnarray}
If~$|\Phi\rangle$ is an eigenstate of~$\hH$ with energy~$E$ then one obtains from Eq.~(\ref{ThhHhPeX})
\begin{equation} \label{ThhHhPe}
 \left(\hH - \hPe \right)\Psi({\bm R}, r,\phi) = \left\{\begin{array}{c} E \\ 0 \end{array} \right\}
 \Psi({\bm R}, r,\phi), \hspace{0.25em}
 \left\{\begin{array}{c} \Psi = \Psi^o \\ \Psi \neq \Psi^o \end{array} \right\}.
\end{equation}
As seen from Eq.~(\ref{ThhHhPe}), even parts of~$\Psi({\bm R}, r,\phi)$ are annihilated
by~$\left(\hH - \hPe \right)$ operator, while odd parts of~$\Psi({\bm R}, r,\phi)$
satisfy the Schrodinger-like equation. For~$\hPo$ one finds
\begin{equation} \label{ThhHhPo}
 \left(\hH - \hPo \right)\Psi({\bm R}, r,\phi) = \left\{\begin{array}{c} E \\ 0 \end{array} \right\}
 \Psi({\bm R}, r,\phi), \hspace{0.25em}
 \left\{\begin{array}{c} \Psi = \Psi^e \\ \Psi \neq \Psi^e \end{array} \right\}.
\end{equation}
Equations~(\ref{ThhHhPe}) and~(\ref{ThhHhPo}) can be treated as alternative definitions
of~$\hPe$ and~$\hPo$ operators.

Consider the functions~$\Psi$,~$\Psi^o$ and~$\Psi^e$ in
Eqs.~(\ref{ThPsie}) and~(\ref{ThPsio}).
Let~$\hT_{\bm a}$ be the translation
operator:~$\hT_{\bm a}w({\bm r}) =w({\bm r}+{\bm a})$.
Then one has~\cite{TranslWiki}
\begin{equation} \label{ThT}
 \hT_{\bm a} = \exp(-i{\bm a} \hat{\bm p}/\hbar),
\end{equation}
where~$\hat{\bm p}=(\hbar/i)\hat{\bm \nabla}_{\bm r}$
is the canonical momentum. Applying the above definition to~$\phi$
coordinate in~$\Psi({\bm R}, r,\phi)$ one obtains
from Eqs.~(\ref{ThPsie}),~(\ref{ThPsio}) and~(\ref{ThT})
\begin{eqnarray}
 \label{ThPsiep}
 \Psi({\bm R}, r,\phi)^{e} = \frac{1}{2} \left( \hI + e^{-i\pi r\hat{p}_{\phi}/\hbar} \right)
 \Psi({\bm R}, r,\phi), \\
 \label{ThPsiop}
 \Psi({\bm R}, r,\phi)^{o} = \frac{1}{2} \left( \hI - e^{-i\pi r\hat{p}_{\phi}/\hbar} \right)
 \Psi({\bm R}, r,\phi),
\end{eqnarray}
where~$\hat{p}_{\phi} = (\hbar/ir)(\partial/\partial \phi)$
is the angular component of the momentum, and~$\hI$ is the unity
operator. We introduce two auxiliary operators
\begin{eqnarray}
 \label{ThAe}
 \hAe = \frac{1}{2} \left( \hI + e^{-i\pi r\hat{p}_{\phi}/\hbar} \right), \\
 \label{ThAo}
 \hAo = \frac{1}{2} \left( \hI - e^{-i\pi r\hat{p}_{\phi}/\hbar} \right).
\end{eqnarray}
Then one has from Eqs.~(\ref{ThPs}),~(\ref{ThPa}), and~(\ref{ThPsiep})--(\ref{ThAo})
\begin{eqnarray}
 \label{TSche}
 \left( \hH -\hPe \right) \Psi({\bm R}, r,\phi) &=& \hH \left[\hAo \Psi({\bm R}, r,\phi) \right], \\
 \label{TScho}
 \left( \hH -\hPo \right) \Psi({\bm R}, r,\phi) &=& \hH \left[\hAe \Psi({\bm R}, r,\phi) \right].
\end{eqnarray}
The meaning of Eq.~(\ref{TSche}) is that the
operator~$\left( \hH -\hPe \right)$, which has only odd states, acting on
a general function~$\Psi({\bm R}, r,\phi)$ gives the same result
as the Hamiltonian~$\hH$ acting on~$\hAo \Psi({\bm R}, r,\phi)$,
which is an odd part of~$\Psi({\bm R}, r,\phi)$.
Solving equations~(\ref{TSche}) and~(\ref{TScho}) for~$\hPe$ and~$\hPo$ one
finds
\begin{eqnarray}
 \label{ThPeA}
 \hPe &=& \frac{1}{2}\hH \left( \hI + e^{-i\pi r\hat{p}_{\phi}/\hbar} \right), \\
 \label{ThPoA}
 \hPo &=& \frac{1}{2}\hH \left( \hI - e^{-i\pi r\hat{p}_{\phi}/\hbar} \right).
\end{eqnarray}
Introducing the total spin:~$\hS=\hs_1 + \hs_2$ one obtains
\begin{equation} \label{ThP}
 \hP = \frac{1}{2}\hH \left(\hI
 + (-1)^{2\hS_z} e^{-i\pi r\hat{p}_{\phi}/\hbar} \right).
\end{equation}
Operators~$\hPe$,~$\hPo$ and~$\hP$ defined in Eqs.~(\ref{ThPeA})--(\ref{ThP}) act
on the function~$\Psi({\bm R}, r,\phi)$.
The representation of~$\hP$, as given in Eqs.~(\ref{ThPeA})--(\ref{ThP}),
exists only in~$2D$, see Discussion.
Inserting~$\hP$ from Eq.~(\ref{ThP}) into Eqs.~(\ref{ThPs}) and~(\ref{ThPa})
one does not obtain the Schrodinger equation for~$\Psi({\bm R}, r,\phi)^{\eta}$
but the differential equations of higher order in~$\hat{p}_{\phi}$, since
\begin{equation} \label{ThExpansion}
 e^{-i\pi r\hat{p}_{\phi}/\hbar} = \sum_{n=0}^{\infty} \frac{1}{n!}
 \left( \frac{-i\pi r\hat{p}_{\phi}}{\hbar} \right)^n.
\end{equation}

The presence of~$\hat{p}_{\phi}$ in the exponents
in Eqs.~(\ref{ThPeA})--(\ref{ThP})
causes a non-locality of~$\hP$ in the position representation.
Using notation:~$|{\bm Q}\rangle = |{\bm R},{\bm r}\rangle$
and~$d{\bm Q}= d^2{\bm R}d^2{\bm r}$ the matrix element
of~$\hPe$ in Eq.~(\ref{ThPeA}) between two~$|{\bm Q}\rangle$ states is
\begin{eqnarray}
 \langle {\bm Q} |\hPe |{\bm Q}'\rangle &=&\int d{\bm Q}''\langle {\bm Q}|\hH|{\bm Q}''\rangle \times
 \nonumber \\
 && \langle {\bm Q''}|\frac{1}{2}\left( \hI + e^{-i\pi r\hat{p}_{\phi}/\hbar} \right)|{\bm Q}'\rangle,
\end{eqnarray}
and similarly for~$\hPo$. In the position representation~$\hH$ in Eq.~(\ref{ThHU})
is a local operator,
so that:~$\langle {\bm Q}|\hH|{\bm Q}''\rangle = \hH_{{\bm Q}{\bm Q}} \delta({\bm Q} - {\bm Q}'')$.
The translation~$e^{-i\pi r\hat{p}_{\phi}/\hbar}$ in Eq.~(\ref{ThPeA}) has nonzero
elements between states~$|{\bm R},r,\phi \rangle$ and~$|{\bm R},r, \phi + \pi\rangle$,
(for~$0 \le \phi < 2\pi$), i.e. between states~$|{\bm R},{\bm r}\rangle$
and~$|{\bm R},-{\bm r}\rangle$. This gives
\begin{eqnarray}
 \label{ThPeR}
 \langle {\bm Q}|\hPe |{\bm Q}'\rangle = \frac{1}{2} \hH_{{\bm Q}{\bm Q}} \delta({\bm R}-{\bm R}')
 \left[ \delta({\bm r}-{\bm r}') + \delta({\bm r}+{\bm r}') \right], \ \ \ \ \ \\
 \label{ThPoR}
 \langle {\bm Q}|\hPo |{\bm Q}'\rangle = \frac{1}{2} \hH_{{\bm Q}{\bm Q}} \delta({\bm R}-{\bm R}')
 \left[ \delta({\bm r}-{\bm r}') - \delta({\bm r}+{\bm r}') \right]. \ \ \ \ \
 \end{eqnarray}
From the above equations one has, see Eq.~(\ref{ThP})
\begin{eqnarray}
\langle {\bm R},{\bm r} |\hP |{\bm R}',{\bm r}'\rangle &=&
 \frac{1}{2} \langle {\bm R},{\bm r} |\hH|{\bm R}',{\bm r}'\rangle\
 \delta\left({\bm R}-{\bm R}'\right) \times \nonumber \\
 \label{ThPR}
 &\times& \left[ \delta({\bm r}-{\bm r}')+ (-1)^{2\hS_z} \delta({\bm r}+{\bm r}') \right]. \ \
\end{eqnarray}
In the position representation one
obtains a non-local equation for the energy levels and wave functions
\begin{equation} \label{ThPsiQ}
 \hH \Psi({\bm Q}) - \int d^2{\bm Q}' \langle {\bm Q}|\hP|{\bm Q}'\rangle
 \Psi({\bm Q}') = \left\{\begin{array}{c} E \\ 0 \end{array} \right\} \Psi({\bm Q}),
\end{equation}
which resembles the Yamaguchi equation~\cite{Yamaguchi1954}.
The second term in Eq.~(\ref{ThPsiQ})
describes a correction to the two-particle Hamiltonian~$\hH$ due to
presence of the Pauli exclusion principle.
Equations~(\ref{ThPR}) and~(\ref{ThPsiQ}) completely describe
the system because they contain {\it all} information necessary
to solve the two-electron problem
including the limitations resulting from the Pauli exclusion principle.
Once~$\hP$ is added to the Hamiltonian,
no additional conditions on multi-electron wave function are needed.

Equations~(\ref{ThPeR})--(\ref{ThPR}) suggest that
in the position representation in~$1D$ and~$3D$ the PEO for two-electron systems
have similar forms. Examples in the Section~\ref{Sec_Examples}
confirm this observation.

\section{Multi-electron systems \label{Sec_MultiE}}

In this section we generalize PEO for systems having more electrons.
The results are more formal and abstract than those
obtained for two-electron systems. Below we provide a definition of PEO
for an arbitrary multi-electron Hamiltonian, but the remaining definition
will relate to three-electron systems.

\subsection{General results}

Let~$\hPi_{ij}$ be the operator exchanging positions of two particles
\begin{equation} \label{MePi}
 \hPi_{ij}|{\bm r}_i{\bm r}_j \rangle = |{\bm r}_j{\bm r}_i \rangle.
\end{equation}
This operator can be expressed as an infinite series of position and
momentum operators. In~$1D$ there is~\cite{Schmid1979}
\begin{equation} \label{MePi1D}
 \hPi_{ij} =\sum_{n=0}^{\infty} \left(\frac{1}{n!} \right)\left(\frac{i}{\hbar}\right)^n
 (\Delta \hat{p}_x)^n(\Delta \hat{r}_x)^n,
\end{equation}
where~$\Delta \hat{r}_x = \hat{r}_{jx} - \hat{r}_{ix}$
and~$\Delta \hat{p}_x = \hat{p}_{jx} - \hat{p}_{ix}$.
The series for~$\hPi_{ij}$ in~$2D$ and~$3D$ are given in
Appendix~\ref{AppA}.

Let~$|\sigma_i\sigma_j\rangle$ be a state of two electron
spins. The operator~$\Sigma_{ij}$ exchanging the spins is,
see Appendix~\ref{AppA}
\begin{equation} \label{MeSig}
 \hSig_{ij} = \frac{1}{2} + 2( {\bm\sigma}_i \cdot {\bm\sigma}_j).
\end{equation}
Then the operator exchanging two electrons is
\begin{equation} \label{Mechi}
 \hchi_{ij}=\hPi_{ij}\hSig_{ij}.
\end{equation}
Let~$|{\rm n}\rangle$ be a state vector of~$k \geq 2$ electrons
\begin{eqnarray} \label{Mermn}
 \langle {\bm r}_1 \sigma_1, \ldots, {\bm r}_k \sigma_k|{\rm n}\rangle
 &=& \Psi({\bm r}_1 \sigma_1, \ldots, {\bm r}_k \sigma_k) \nonumber \\
 &=& \Psi(1,\ldots,k).
\end{eqnarray}
Then we define PEO as
\begin{equation} \label{MeDefP}
 \left(\hH - \chP \right) \Psi(1,\ldots,k) =
 \hH \left(\prod_{i=1,j>i}^k\hchi_{ij} \Psi(1,\ldots,k) \right).
\end{equation}
The physical meaning of~$\chP$ is that
the operator~$\left(\hH - \chP \right)$ acting on unrestricted
function~$\Psi(1,\ldots,k)$
gives the same result as the Hamiltonian~$\hH$ acting on a function
that is antisymmetric with respect to
exchange of all pairs of electrons. Note that~$\chP$ in Eq.~(\ref{MeDefP}) is defined
in a different way than~$\hPe$ and~$\hPo$ in Eqs.~(\ref{ThPs}) and~(\ref{ThPa}),
see Discussion. By solving Eq.~(\ref{MeDefP}) one obtains
\begin{equation} \label{MechP}
 \chP \Psi(1,\ldots,k) = \left[\hH \left(\hI - \prod_{i=1,j>i}^k\hchi_{ij} \right)\right]
 \Psi(1,\ldots,k).
\end{equation}
Equation~(\ref{MechP}) generalizes Eqs.~(\ref{TSche}) and~(\ref{TScho})
for multi-electron case.
Let~$\{|{\rm n}\rangle\}$ and~$\{E_{\rm n}\}$ be the complete sets of
states and energies of multi-electron Hamiltonian~$\hH$, respectively.
Let~$\{|{\rm n}^a\rangle\}$ be a subset of~$\{|{\rm n}\rangle\}$
including states antisymmetric with respect to exchange of all pairs
of electrons~$({\bm r}_i \sigma_i) \Leftrightarrow ({\bm r}_j\sigma_j)$
for~$1 \leq i, j \leq k$. Then PEO is
\begin{equation} \label{MeRes}
 \chP = \sum_{\rm n} E_{\rm n} \Big( |{\rm n} \rangle \langle {\rm n}| -
 |{\rm n}^a \rangle \langle {\rm n}^a| \Big) =
 \sum_{{\rm n} \notin \{{\rm n}^a \}} E_{\rm n} |{\rm n} \rangle \langle {\rm n} |.
\end{equation}
As seen from Eq.~(\ref{MeRes}), spectral resolution of PEO
includes all states of~$\hH$ except those that are antisymmetric with respect
to exchange of all pairs of electrons.
Equation~(\ref{MeRes}) generalizes Eqs.~(\ref{ThPe}) and~(\ref{ThPo}) for
multi-electron systems. To find the analogue of Eqs.~(\ref{ThhHhPe})
and~(\ref{ThhHhPo}) we insert Eqs.~(\ref{Mermn}) and~(\ref{MeRes})
into Eq.~(\ref{MeDefP}) and obtain
\begin{equation} \label{MehHchP}
 \left( \hH - \chP \right) |{\rm n}\rangle =
 \left[\begin{array}{c} E_{\rm n} \\ 0 \end{array} \right] |{\rm n}\rangle,
\end{equation}
where the upper identity holds
for~$|{\rm n}\rangle \in \{|{\rm n}^a\rangle\}$ and the lower one
for~$|{\rm n}\rangle \notin \{|{\rm n}^a\rangle\}$.
As follows from Eq.~(\ref{MehHchP}), operator~$\left( \hH - \chP \right)$
annihilates states~$|{\rm n}\rangle$ of improper symmetry with respect to exchange
of all pairs of electrons, while states of proper symmetry satisfy the Schrodinger-like
equation.

\subsection{Approximations}

Since it is difficult to obtain the exact form of PEO for multi-electron
systems we describe here several possible approximations of~$\chP$.
The natural approximation to~$\chP$ is truncation of infinite series
in Eqs.~(\ref{MePi1D}),~(\ref{AppA_hPi3}) and~(\ref{AppA_hPi2})
to large but finite number of terms. Then one obtains
a high-order differential equation that can be solved by standard methods.
Attention should be paid to the domain of series convergence in
Eqs.~(\ref{MePi1D}),~(\ref{AppA_hPi3}) and~(\ref{AppA_hPi2}).
An alternative expression for permutation operator is
given in Ref.~\cite{Grau1981}.

In the second approach one may approximate in Eq.~(\ref{MechP}) the exact
operator~$\prod_{i=1,j>i}^k\hchi_{ij}$ by a simpler one using results
from the previous section. Consider the four-electron case,
the function~$\Psi({\bm r}_1,{\bm r}_2, {\bm r}_3, {\bm r}_4)$,
and disregard electrons spins. Let us introduce two pairs of
center-of-mass and relative-motion coordinates, see Eq.~(\ref{ThRr}). Then
one obtains a set of functions in the form
\begin{equation} \label{MaPsi_ijkl}
 \Psi_{ij,kl}({\bm R}_{ij},{\bm r}_{ij}, {\bm R}_{kl}, {\bm r}_{kl}), \hspace{0.5em} 1
 \leq i,j,k,l \leq 4,
\end{equation}
and each of them satisfies Eq.~(\ref{ThPsiQ}) with PEO similar to that
in Eq.~(\ref{ThPR}) for appropriate pairs of coordinates. Each of functions
in Eq.~(\ref{MaPsi_ijkl}) is symmetric
or antisymmetric in two pairs of variables (instead of all pairs), but having all
set of function~$\Psi_{ij,kl}$ one may approximate the true function~$\Psi$.

In the third approximation one replaces the exact Hamiltonian~$\hH$ entering
to PEO in Eq.~(\ref{MehHchP}) by a simpler one~$\hH_0$, as e.g.
that of~$k \geq 2$ free electrons in a harmonic potential.
Let~$|\Psi \rangle$ be~$k$-electron state
and~$\chP_0$ be PEO corresponding to~$\hH_0$. Then one has
\begin{equation}
 \left(\hH - \chP\right) |\Psi\rangle \simeq \left(\hH - \chP_0 \right) |\Psi \rangle.
\end{equation}
Using Eq.~(\ref{MeRes}) one finds
\begin{equation} \label{Man0a}
 \left(\hH - \chP\right) |\Psi\rangle \simeq \hH |\Psi\rangle - \lambda\left(\sum_{{\rm n}_0^a}
 E_{{\rm n}_0^a}
 |{\rm n}_0^a \rangle \langle {\rm n}_0^a|\right) |\Psi\rangle = E |\Psi\rangle,
\end{equation}
where~$\lambda$ is a parameter,~$|{\rm n}_0^a \rangle$ are
antisymmetric states of~$\hH_0$ with respect to exchange of all
pairs of electrons and~$E_{{\rm n}_0^a}$ are the corresponding energies.
The summation in Eq.~(\ref{Man0a}) is restricted to a finite
number of states.
The presence of~$\lambda$ in Eq.~(\ref{Man0a}) allows one to
switch on the approximate PEO to the Schrodinger equation. If the obtained
function~$\Psi$ has proper symmetry with respect to exchange of all pairs of
electrons then both~$\Psi$ and the corresponding energy~$E$ weakly
depend on~$\lambda$ since in this case the second tern in Eq.~(\ref{Man0a})
vanishes or is small. If the calculated function~$\Psi$ has improper
symmetry, then both~$\Psi$ and~$E$ strongly depend on~$\lambda$ because
in this case the second term in Eq.~(\ref{Man0a}) is large and it
strongly influences~$\Psi$ and~$E$.
The described approach gives a practical method of finding multi-electron
states having proper symmetry with respect to exchange of all
pairs of electrons. Example of such calculations for Hooke's atom
is shown in the next section.

A possible generalization of Eq.~(\ref{Man0a}) is to
treat the second term in this equation as a kernel
operator that ensures the antisymmetry of the resulting function~$\Psi$ for
some set of states, e.g., low-energy ones.
Let~$|\bm Q\rangle=|{\bm r}_1,\ldots, {\bm r}_k\rangle$
and~$\Psi({\bm Q}) = \Psi({\bm r}_1,\ldots, {\bm r}_k)$.
Then one has from Eq.~(\ref{Man0a})
\begin{eqnarray}
 \langle {\bm Q}|\hH - \chP| \Psi \rangle &\simeq&
 \hH \Psi({\bm Q}) - \lambda \int \hK({\bm Q}', {\bm Q}) \Psi({\bm Q}') d{\bm Q}'
 \nonumber \\
 \label{MaKer}
 &=& E\Psi({\bm Q}).
\end{eqnarray}
Comparing Eqs.~(\ref{Man0a}) and~(\ref{MaKer}) one finds
\begin{equation}
 \label{MaKQQ}
 \hK({\bm Q}, {\bm Q}') = \sum_{{\rm n}_0^a} E_{{\rm n}_0^a}
 \langle {\bm Q}|{\rm n}_0^a \rangle \langle {\rm n}_0^a| {\bm Q}' \rangle.
\end{equation}
The idea of kernel approach is that~$\hK({\bm Q}, {\bm Q}')$
in Eq.~(\ref{MaKer}) can be any mathematical operator without physical meaning.
As an example, when in Eq.~(\ref{MaKQQ}) one replaces energies~$E_{{\rm n}_0^a}$
by a constant value~${\cal E}_c$
and limits the summation to~${\rm n}_{max}$ terms, one obtains simpler expression
\begin{equation}
 \label{MaK1QQ}
 \hK_1({\bm Q}, {\bm Q}') = {\cal E}_c \sum_{{\rm n}_0^a}^{{\rm n}_{max}}
 \langle {\bm Q}|{\rm n}_0^a \rangle \langle {\rm n}_0^a| {\bm Q'} \rangle,
\end{equation}
that also selects states having proper symmetry with respect to exchange of all
pairs of electrons. However, the kernel in Eq.~(\ref{MaK1QQ}) works correctly
only for states having similar energies to those corresponding to
functions~$\langle {\bm Q}|{\rm n}_0^a\rangle$ in Eq.~(\ref{MaK1QQ}).
The example of kernel approach to Hooke's atom is given in the next section.

Finally, we discuss approximation in which one calculates the expected value
of~$\left(\hH -\chP \right)$ over a trial function~$|\Phi^a\rangle$ that is
already antisymmetric with respect to exchange of all pairs of electrons.
Assuming that~$\langle \Phi^a|\Phi^a \rangle=1$ one obtains from Eq.~(\ref{MeRes})
\begin{equation} \label{MaPEq0}
 \langle \Phi^a\left| \chP \right| \Phi^a \rangle = 0,
\end{equation}
since in this case the trial functions~$|\Phi^a\rangle$ is a linear combination of
states~$|{\rm n}^a\rangle$ that are antisymmetric with respect to exchange of all
pairs of electrons, while~$\chP$ does not include these states in its spectral
resolution, see Eq.~(\ref{MeRes}). Then
\begin{equation} \label{MaPSl}
 \langle \Phi^a\left| \hH-\chP \right| \Phi^a \rangle \equiv
 \langle \Phi^a\left| \hH\right| \Phi^a \rangle = E_a,
\end{equation}
where~$E_a$ is approximated energy.
A practical consequence of Eqs.~(\ref{MaPEq0}) and~(\ref{MaPSl})
is that, when one calculates variationally energies and states of
multi-electron system with trial function in the form of Slater
determinant, then PEO identically vanishes and there is no need to
introduce it to calculations.

\section{Examples of~$\hP$ operators for Hooke's atom \label{Sec_Examples}}

Here we show two examples of~$\hP$ for two-electron systems
and rederive analytically or numerically the results of
Eqs.~(\ref{ThPeR})--(\ref{ThPR})
by explicit summations over even or odd states of the Hamiltonian spectrum,
see Eqs.~(\ref{ThPe}) and~(\ref{ThPo}).

We consider first the Hooke's atom in~$2D$
whose Hamiltonian is given in Eq.~(\ref{ThHU}) with~$U({\bm r}_i)= kr_i^2/2$
and~$i=1,2$, where~$k>0$ is the harmonic potential
strength~\cite{Kais1989,Taut1993,Taut1994,ONeil2003,HookesWiki}.
Then~$\hH$ separates into two parts~$\hH_{\bm R}$ and~$\hH_{\bm r}$
depending on~${\bm R}$ and~${\bm r}$, respectively.
The eigenfunctions of~$\hH$ are~$\Psi({\bm R},{\bm r})=F({\bm R}) f({\bm r})$,
where~$F({\bm R})$ and~$f({\bm r})$ satisfy equations
\begin{eqnarray}
 \left(-\frac{1}{4} {\bm \nabla}_{\bm R}^2 +kR^2 \right)F({\bm R}) &=& E_R F({\bm R}), \label{ExF} \\
 \left(- {\bm \nabla}_{\bm r}^2 + \frac{1}{r}
 + \frac{1}{4} kr^2 \right) f_{m,n}({\bm r}) &=& E_{m,n} f_{m,n}({\bm r}), \label{Exf}
\end{eqnarray}
where~$E_{m,n}$ is the energy of~$n$-th state with the angular momentum number~$m$.
The center-of-mass motion, as given in Eq.~(\ref{ExF}),
is described by~$2D$ harmonic oscillator.
For the relative motion in Eq.~(\ref{Exf})
we set:~$f_{m,n}({\bm r})= g_{m,n}(r)e^{im\phi}/\sqrt{2\pi}$,
where~$g_{m,n}(r)$ are solutions of
\begin{equation}
 \label{ExHg}
 \left( -\frac{d^2}{dr^2} - \frac{1}{r}\frac{d}{dr} +\frac{m^2}{r^2} + \frac{1}{r}
 + \frac{k}{4}r^2 \right) g_{m,n}(r) = E_{m,n} g_{m,n}(r).
\end{equation}

Consider the operator~$\hPo$ in Eq.~(\ref{ThPo}).
Since~$\hH_{\bm R}$ in Eq.~(\ref{ExF}) is not affected by~$\hPo$
we concentrate on~$\hH_{\bm r}$. Let~$|m,n\rangle$ be an eigenstate of Eq.~(\ref{Exf}),
and~$\langle{\bm r}|m,n\rangle = f({\bm r})$. Then one has
\begin{eqnarray}
 \hPo &=& \sum_{m=-\infty}^{\infty} \sum_{n=1}^{\infty} E_{2m+1,n} |2m+1,n\rangle\langle 2m+1,n| \nonumber \\
 \label{ExPo1}
 &=& \hH \left(\sum_{m=-\infty}^{\infty} \sum_{n=1}^{\infty} |2m+1,n\rangle\langle 2m+1,n| \right).
\end{eqnarray}
In the position representation there is
\begin{eqnarray}
 \langle {\bm r}|\hPo|{\bm r}'\rangle &=& \frac{1}{2\pi} \int d^2 {\bm r}''
 \langle {\bm r}|\hH|{\bm r}''\rangle
 \sum_{m=-\infty}^{\infty} e^{i(2m+1)(\phi''-\phi')} \times \nonumber \\
 \label{ExPoH1}
 && \sum_{n=1}^{\infty} g_{2m+1,n}(r'')^*g_{2m+1,n}(r').
\end{eqnarray}
We first calculate the sum over~$n$.
The functions~$g_{m,n}(r)$ are normalized using the weight
function~$w_g(r)=r$. Consider
functions~$h_{m,n}(r) = \sqrt{r}g_{m,n}(r)$
normalized using the weight function~$w_h(r)=1$.
They are eigenfunctions of equation, see Eq.~(\ref{ExHg})
\begin{equation}
 \label{ExHh}
 \left( -\frac{d^2}{dr^2} +\frac{m^2-1/4}{r^2} + \frac{1}{r}
 + \frac{k}{4}r^2 \right) h_{m,n}(r) = E_{m,n} h_{m,n}(r).
\end{equation}
For fixed~$m$, functions~$h_{m,n}(r)$ form a complete
set of states of the Hermitian operator in Eq.~(\ref{ExHg}), so there is
\begin{equation}
 \label{ExHhs}
 \sum_{n=1}^{\infty} h_{2m+1,n}(r'')^*h_{2m+1,n}(r') = \delta(r'-r''),
\end{equation}
which gives
\begin{equation}
 \label{ExHgs}
 \sum_{n=1}^{\infty} g_{2m+1,n}(r'')^*g_{2m+1,n}(r') = \frac{\delta(r'-r'')}{r''},
\end{equation}
and the result of summation over~$n$ does not depend on~$m$.
Consider now the sum over~$m$ in Eq.~(\ref{ExPoH1}).
Let~$\xi=\phi''-\phi'$. Then one has
\begin{equation}
 \label{ExSumm}
 \frac{1}{2\pi}\sum_{m=-\infty}^{\infty}\!\! e^{i(2m+1)\xi}=
 \frac{e^{i\xi}}{2\pi}\!\! \sum_{m=-\infty}^{\infty}\!\! e^{im(2\xi)} =
 \frac{e^{i\xi}}{2} \delta(\xi - N\pi),
\end{equation}
which gives:~$(\phi''-\phi')=0$ or~$(\phi''-\phi')=\pi$,
since~$(\phi'-\phi'')\in [0,2\pi)$.
In Eq.~(\ref{ExSumm}) we used
identity:~$\sum_{m=-\infty}^{\infty} e^{im\xi}= 2\pi\delta(\xi-2N\pi)$
with~$N$ integer. Then one obtains
\begin{eqnarray}
\label{ExPo03}
 \langle {\bm r}|\hPo |{\bm r}'\rangle=
 \int d^2 {\bm r}'' \langle {\bm r}|\hH|{\bm r}''\rangle \times\nonumber \\
 \left[\frac{1}{2}\delta(\phi''-\phi') + \frac{e^{i\pi}}{2} \delta(\phi''-\phi' + \pi) \right]\!\!
 \left[\frac{1}{r''} \delta(r''-r') \right]. \ \ \
\end{eqnarray}
There is~$\langle {\bm r}|\hH|{\bm r}''\rangle = \delta({\bm r}-{\bm r}'')$
since the Hamiltonian is a local operator.
Using the identity:~$\delta({\bm r}-{\bm r}') = (1/r)\delta(r-r')\delta(\phi-\phi')$
for~$2D$ delta function one obtains Eq.~(\ref{ThPoR}). The generalization of this approach
to~$1D$ and~$3D$ Hooke's atoms is straightforward.

In the second example we calculate numerically operator~$\hPo$ in a system in which
the functions~$f(r,\phi)$ do not separate into products of two one-dimensional functions.
Consider the model similar to the Hooke's atom in Eq.~(\ref{ExHg})
but with non-radial external potential. Its Hamiltonian is
given by Eq.~(\ref{ThHU}) with~$U({\bm r}_i)= k_xx_i^2/2 + k_yy_i^2/2$ and~$i=1,2$.
The potential strengths~$k_x, k_y > 0$.
Introducing center-of-mass and relative motion coordinates one obtains
\begin{eqnarray}
 \left(-\frac{1}{4} {\bm \nabla}_{\bm R}^2 +k_xX^2 + k_yY^2 \right)F({\bm R})
 &=& E_R F({\bm R}), \ \ \ \ \label{ExEqF} \\
 \left(-\frac{\partial^2}{\partial r^2} - \frac{1}{r}\frac{\partial}{\partial r}
 - \frac{1}{r^2}\frac{\partial^2}{\partial \phi^2} + \frac{1}{r} + \right. \nonumber \\
 \left. + \frac{1}{4} k_yr^2 + qr^2\cos(\phi)^2\right) f({\bm r}) &=& E_r f({\bm r}). \ \ \ \ \label{ExEqf}
\end{eqnarray}
where~$q=k_x-k_y$ characterizes anisotropy of the external potential.
To find~$\hPo$ we expand functions~$f(r,\phi)$ in Eq.~(\ref{ExEqf})
into the set of eigenstates~$g_{m,n}(r)e^{im\phi}/\sqrt{2\pi}$
of the Hooke's atom, see Eq.~(\ref{ExHg})
\begin{equation}
 f(r,\phi) = \sum_{m=-m_{max}}^{m_{max}}\sum_{n=1}^{n_{max}} b_{m,n}g_{m,n}(r)e^{im\phi},
\end{equation}
where~$b_{m,n}$ are the expansion coefficients,~$m_{max}=16$ and~$n_{max}\simeq 250$.
The presence of Hooke's atom functions in Eq.~(\ref{ExHg}) ensures orthogonality
of the basis. We used~$8054$ basis functions~$g_{m,n}(r)$, which are calculated by
the shooting method, see Appendix~\ref{AppB}. We introduce a mapping:~$(m,n) \rightarrow i$
which labels the basis functions~$g_{m,n}(r)$ with a single index~$i$.

\begin{figure} \includegraphics[width=8cm,height=12cm]{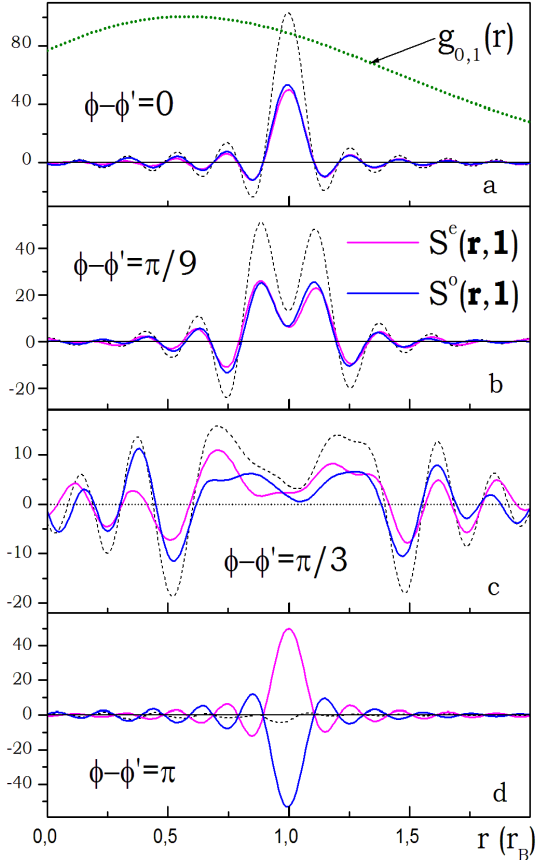}
 \caption{Dimensionless sums~$\cSo({\bm r},{\bm 1})$ and~$\cSe({\bm r},{\bm 1})$
 given in Eq.~(\ref{ExSeta}) calculated numerically for nonsymmetric~$2D$
 Hooke's atom in Eq.~(\ref{ExEqf}) for several
 values of relative phases~$(\phi-\phi')$. The dashed lines represent
 sums~$\cSo({\bm r},{\bm 1}) +\cSe({\bm r},{\bm 1})$ approximating
 delta function~$\delta({\bm r}-{\bm 1})$. In panel a) the dotted line
 indicates ground-state function~$g_{0,1}(r)$
 of~$2D$ Hooke's atom in Eq.~(\ref{Exf}).} \label{Fig1} \end{figure}

The eigenenergies and eigenstates of the Hamiltonian in Eq.~(\ref{ExEqf}) are
obtained by solving the problem of finite-size
matrix:~$\sum_{i'}H_{ii'} a_{i'} = E a_i$, where~$a_i$ are uniquely
obtained from~$b_{m,n}$ by the mapping:~$i \rightarrow (m,n)$.
Using the inverse mapping~$(m,n) \rightarrow i$ one has
\begin{equation}
 H_{ii'} = E_{i}\delta_{i,i'} + qc_{m,m'}\int_0^{\infty}
 \left[r^2 g_{m,n}(r) g_{m',n'}(r) \right] r dr,
\end{equation}
where for fixed~$m$ the functions~$g_{m,n}$ are
normalized:~$\int_0^{\infty}g_{m,n}(r)g_{m,n'}(r)r dr = \delta_{n,n'}$. The selection
rules for~$\phi$ integrals are
\begin{eqnarray}
 c_{m,m'} &=& \frac{1}{2\pi} \int_0^{2\pi} e^{i(m-m')\phi}\cos(\phi)^2 d\phi \nonumber \\
 &=& \frac{1}{2}\delta_{m,m'} + \frac{1}{4}\delta_{m,m'\pm 2}.
\end{eqnarray}
The nonzero elements of~$H_{ii'}$ are those with~$m'=m$ and~$m'=m\pm 2$.
Let~$\{f_l^e(r,\phi)\}$ be a set of states of~$H_{ii'}$ obtained
from even functions~$g_{2m,n}(r)e^{(2m)i\phi}/\sqrt{2\pi}$,
and~$\{f_l^o(r,\phi)\}$ be a set of states of~$H_{ii'}$
obtained from odd functions~$g_{2m+1,n}(r)e^{(2m+1)i\phi}/\sqrt{2\pi}$. Then
\begin{equation}
 \label{ExPeta}
 \hP^{\eta}({\bm r},{\bm r}') = \hH({\bm r},{\bm r}) \ \cSeta({\bm r},{\bm r}').
\end{equation}
where
\begin{equation}
 \label{ExSeta}
 \cSeta({\bm r},{\bm r}') = \sum_l f_l^{\eta}(r,\phi)^*f_l^{\eta}(r',\phi'),
\end{equation}
and~$\eta \in \{e,o\}$. Note that for~$l \rightarrow \infty$ there
is:~$\cSe({\bm r},{\bm r}') + \cSo({\bm r},{\bm r}') \rightarrow \delta({\bm r},{\bm r}')$.
In our calculations we take~$4146$~$f_l^e(r,\phi)$ functions and~$3908$~$f_l^o(r,\phi)$
functions, respectively.

\begin{figure} \includegraphics[width=8cm,height=8cm]{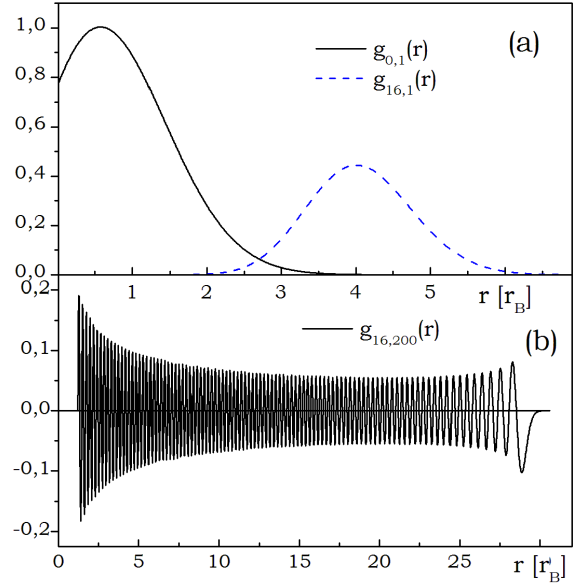}
 \caption{Functions~$g_{m,n}(r)$ of 2D Hooke's atom given in Eq.~(\ref{ExHg})
 calculated numerically for three~$m,n$ values. Function~$g_{0,1}(r)$
 corresponds to the ground state of the system.} \label{Fig2} \end{figure}

In Figure~1 we plot the sums~$\cSeta({\bm r},{\bm r}')$ in Eq.~(\ref{ExSeta})
for~${\bm r}'={\bm 1}$ and several~$(\phi-\phi')$ values,
where~${\bm 1}$ is a unit vector in arbitrary direction.
In our calculations we take~$k_y=4$ and~$k_x=9.61$, which
gives~$q=1.4025$, see Eq.~(\ref{Exf}).
In Figure~1a there is~$(\phi-\phi')=0$ and
both sums~$\cSeta({\bm r},{\bm 1})$ tend to~$\delta(r-1)$,
where~$r=|{\bm r}|$. We also plot the un-normalized function~$g_{0,1}(r)$.
It is seen that~$\cSeta$ are more localized than~$g_{0,1}(r)$ which justifies
treating~$\cSeta({\bm r},{\bm 1})$ as approximations of~$\delta(r-1)$ function.

By increasing~$(\phi-\phi')$ in Figures~1b and~1c
the sums~$\cSeta({\bm r},{\bm 1})$ gradually decrease,
but they do not vanish because they are truncated to finite number of terms.
For~$(\phi-\phi')=\pi$ in Figure~1d, the sum~$\cSe({\bm r},{\bm 1})$ tends to~$\delta(r-1)$,
while the sum~$\cSo({\bm r},{\bm 1})$ tends to~$-\delta(r-1)$,
so their sum practically cancels out (dotted line).
The above results obtained numerically in Figure~1 for a non-separable function~$f(r,\phi)$
illustrate general formulas in Eqs.~(\ref{ThPeR})--(\ref{ThPR}).

We emphasize two approximations related to Figure~1. First, the summations over
angular states are limited to~$0 \le m \le 16$, and the results may be incomplete
because we omitted basis functions with higher~$m$. Second, for fixed~$m$ we
take~$n \simeq 250$ radial functions~$g_{m,b}(r)$ and claim that
they are sufficient to approximate combinations of delta functions
in Eqs.~(\ref{ThPeR}) and~(\ref{ThPoR}).
Both issues are clarified in Figures~2 and~3.

\begin{figure} \includegraphics[width=8cm,height=8cm]{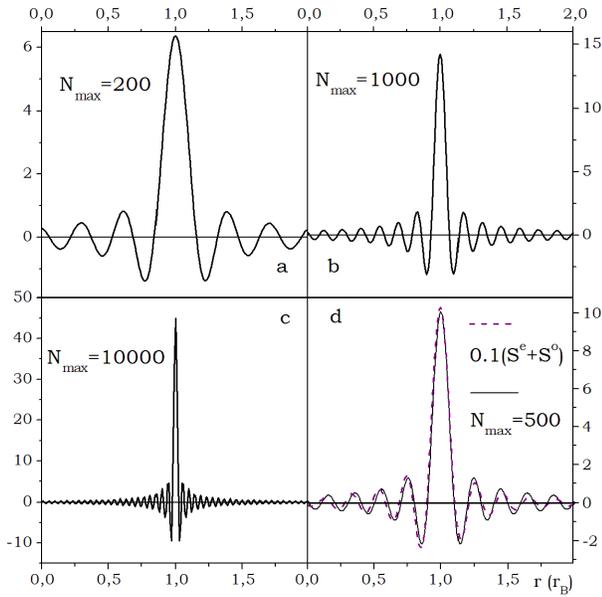}
 \caption{a), b), c) Dimensionless sums~$S(x,x')$ given in Eq.~(\ref{ExS})
 for~$1D$ harmonic oscillator functions calculated numerically for four~$N_{max}$
 values and~$x'=1$. d) Sum~$S(x,1)$ calculated for~$N_{max}=500$ compared
 with re-scaled sum~$\cSe({\bm r},{\bm 1}) + \cSo({\bm r},{\bm 1})$ defined
 in Eq.~(\ref{ExSeta}) and shown in Figure~1a, dashed line.
 The scaling factor is~$c=0.1$.} \label{Fig3}
\end{figure}

In Figure~2 we show
normalized functions~$g_{0,1}(r)$ (ground state),~$g_{16,1}(r)$, and~$g_{16,200}(r)$.
As seen from Figures~2b and~2c, functions having~$m=16$ practically vanish at~$r=1$
and they give negligible contributions to~$\cSeta({\bm r},{\bm 1})$ for~$0 \le r \le 2$,
see Eq.~(\ref{ExSeta}). This result confirms the validity of
truncating the summation over~$m$ states to~$m \le 16$ in Figure~1.
Selecting larger~$r'$ and~$r$ one has to include states with larger~$m$.

To show that finite sums~$\cSeta$ in Figure~1 approximate the combinations
of delta functions we consider the set of functions~$\{\psi_{n}(x)\}$
being states of the one-dimensional harmonic oscillator with
the potential~$U(x) = x^2$. Let
\begin{equation} \label{ExS}
 S(x,x') = \sum_{n}^{N_{max}} \psi_{n}(x)\psi_{n}(x') \rightarrow \delta(x-x').
\end{equation}
We calculate~$S(x,x')$ numerically using the recursion relation~\cite{HermiteWiki}:
$\sqrt{\frac{n+1}{2}} \psi_{n+1}(x) = x\psi_n(x) - \sqrt{\frac{n}{2}}\psi_{n-1}(x)$
with the initial conditions:~$\psi_0(x)= \pi^{-1/4}\exp(-x^2/2)$ and~$\psi_{-1}(x)=0$.
In Figure~3 we show~$S(x,1)$ for several values of~$N_{max}$.
As seen in Figures~3a,~3b, and~3c, when increasing~$N_{max}$ the sums~$S(x,1)$
tend to~$\delta(x-1)$.
In Figure~3d we compare the sum~$S(x,1)$ for~$N_{max}=500$
with the re-scaled sum~$\cSe({\bm r},{\bm 1}) + \cSo({\bm r},{\bm 1})$ for~$(\phi-\phi')=0$
shown in Figure~1a. Both curves are close to each other up to a scaling factor~$c=0.1$,
which confirms the delta-like character of the curves shown in Figure~1.

As the third example we calculate the states and
energies of the symmetric~$2D$ Hooke's atom described in Eq.~(\ref{ExHg})
with use of Eqs.~(\ref{Man0a}) and~(\ref{MaKer}). We analyze
odd states of~$\hH$, so we apply~$\hPo$ operator, see Eq.~(\ref{ThPo}).
In the position representation~$|{\bm Q}\rangle =|{\bm R},{\bm r}\rangle$
equations~(\ref{Man0a}) and~(\ref{MaKer}) read
\begin{eqnarray}
 \hH \Psi({\bm Q}) - \left( \hH_0 - \hPo_0 \right) \Psi({\bm Q})
 = \nonumber \\ \hH \Psi({\bm Q}) -\lambda \sum_{{\rm n}_0^e} E_{{\rm n}_0^e}
 \langle {\bm Q}|{\rm n}_0^e\rangle
 \int \langle {\rm n}_0^e| {\bm Q}' \rangle
 \Psi({\bm Q}') d^2 {\bm Q}' \nonumber \\
 \label{ExPsiQ}
 = E \Psi({\bm Q}), \ \ \
\end{eqnarray}
and~${\bm R}, {\bm r}$ are the center-of-mass
and relative-motion coordinates, respectively. The superscript~$e$ in
Eq.~(\ref{ExPsiQ}) denotes even states and energies of~$\hH_0$,
since the odd ones were eliminated by~$\hPo_0$. Let
\begin{eqnarray}
 \Psi({\bm R}, {\bm r}) &=& \frac{1}{\sqrt{2\pi}}F({\bm R})g_{m,n}(r)e^{im\phi}, \\
 \label{Expsi}
 \langle {\bm R},{\bm r}|{\rm n}_0^a \rangle &=&
 \frac{1}{\sqrt{2\pi}}F({\bm R})\psi_{2j,l}(r)e^{2ij\phi},
\end{eqnarray}
where~$F({\bm R})$ satisfies Eq.~(\ref{ExF}),~$g_{m,n}(r)$ is solution
of Eq.~(\ref{ExHg}),~$\psi_{2j,l}(r)$ and~$\epsilon_{2j,l}$
are functions and energies of~$2D$ harmonic
oscillator, respectively,~$m, j$ describe angular
momentum and~$n,l$ label the discrete states.
Functions~$\psi({\bm r}) =\psi_{2j,l}(r)e^{2ij\phi}$ in Eq.~(\ref{Expsi})
are even:~$\psi({\bm r}) = \psi(-{\bm r})$.
We approximate~$\chP_0$ in Eq.~(\ref{ExPsiQ}) by restricting summations
to few low-energy states:~$j=0, \pm 1$ and~$n=0,1,2$.
For given~$m$ and~$n$ one has from Eq.~(\ref{ExPsiQ})
\begin{eqnarray}
 \hH_r g_{m,n}(r)\frac{e^{im\phi}}{\sqrt{2\pi}} -
 \lambda \sum_{l=0}^2 \sum_{j=-1}^1 \epsilon_{2j,l}\phi_{2j,l}(r)
 \frac{ e^{2ij\phi}}{\sqrt{2\pi}} \times \nonumber \\
 \int_0^{\infty}\phi_{2j,l}(r') g_{m,n}(r')r'dr'
 \int_0^{2\pi} \frac{e^{i(m-2j)\phi'}}{2\pi} d\phi' \nonumber \\
 \label{Exgpsi}
 =E_{m,n} g_{m,n}(r) \frac{e^{im\phi}}{\sqrt{2\pi}}, \ \ \ \ \
\end{eqnarray}
where~$\hH_r$ is defined in Eq.~(\ref{ExHg}) and we
used~$\int |F({\bm R}')d^2{\bm R}'|^2=1$.
The kernel corresponding to Eq.~(\ref{Exgpsi}) is,
see Eqs.~(\ref{MaKQQ}) and~(\ref{MaK1QQ})
\begin{equation}
 \label{ExKer}
 \hK({\bm r}, {\bm r}') = \sum_{l=0}^2 \sum_{j=-1}^1
 \epsilon_{2j,l}\phi_{2j,l}(r) \phi_{2j,l}(r') \frac{e^{2ij(\phi-\phi')}}{2\pi}.
\end{equation}

Now we discuss solutions
of Eq.~(\ref{Exgpsi}) for various values of~$m$ and we analyze three
cases:~$m=\pm 1$,~$m=0, \pm 2$ and~$|m| > 2$.
Consider first two odd states with~$m=\pm 1$. Since the second integral
in Eq.~(\ref{Exgpsi}) vanishes for~$m=\pm 1$ one obtains
\begin{equation} \label{Exgo}
 \hH_r g_{m,n}(r) = E_{m,n} g_{m,n}(r),
\end{equation}
i.e. Eq.~(\ref{ExHg}). The solutions of Eq.~(\ref{Exgo}) do not depend
on~$\lambda$. If in Eq.~(\ref{Exgpsi}) one uses the kernel~$\hK_1({\bm r}, {\bm r}')$
of the form, see Eq.~(\ref{MaK1QQ})
\begin{equation}
 \label{ExKer1}
 \hK_1({\bm r}, {\bm r}') = {\cal E}_c \sum_{j=-1}^1
 \phi_{2j,0}(r) \phi_{2j,0}(r') \frac{e^{2ij(\phi-\phi')}}{2\pi},
\end{equation}
then for~$g_{m,n}(r)$ one also obtains equation~(\ref{Exgo}).
In Eq.~(\ref{ExKer1}) the sum over~$l$ is limited to a single term
with~$l=0$ and~${\cal E}_c$ is an arbitrary energy.

Consider now three even states with~$m=0,\pm 2$. Then the sum over~$j$
in Eq.~(\ref{Exgpsi}) reduces to a single term with~$2j=m$ and one has
\begin{eqnarray}
 \hH_r g_{m,n}(r)-\lambda \sum_{l=0}^2 \epsilon_{m,l}\phi_{m,l}(r)
 \int_0^{\infty}\phi_{m,l}(r') g_{m,n}(r')r'dr' \nonumber \\
 \label{Exge}
 = E_{m,n} g_{m,n}(r). \ \ \ \ \ \ \
\end{eqnarray}
Equation~(\ref{Exge}) is differential-integral equation for
unknown function~$g_{m,n}(r)$, and it resembles Eq.~(\ref{ThPsiQ}).
In Eq.~(\ref{Exge}) the function~$g_{m,n}(r)$ does not vanish and it depends
on~$\lambda$. This also occurs when in Eq.~(\ref{Exgpsi}) one
replaces the kernel~$\hK({\bm r}, {\bm r}')$
by~$\hK_1({\bm r}, {\bm r}')$ in Eq.~(\ref{ExKer1}).

Consider now the exact operator~$\chP$ instead of~$\chP_0$. Then
we set in Eq.~(\ref{Exge})~$\psi_{j,l}(r) \rightarrow g_{m,n}(r)$
and~$\epsilon_{j,l} \rightarrow E_{m,n}$. For~$m=0,\pm 2$ one has
\begin{equation}
 \label{Exge0}
 \hH_r g_{m,n}(r)-\lambda E_{m,n} g_{m,n}(r) = E_{m,n} g_{m,n}(r).
\end{equation}
For~$\lambda=1$ the left-hand-side of Eq.~(\ref{Exge0}) vanishes,
which gives~$g_{m,n}(r) \equiv 0$, as expected from Eq.~(\ref{MehHchP})
for the exact~$\chP$ operator.

Finally, for~$|m| > 2$ one obtains Eq.~(\ref{Exgo})
both for odd and even~$m$, since the approximate PEO in Eq.~(\ref{Exgpsi}) contains
only states with angular momenta~$|m| \leq 2$. This also occurs
for kernel~$\hK_1({\bm r}, {\bm r}')$ in Eq.~(\ref{ExKer1}).

From the above results we reach the following conclusions.
First, by properly chosen set of~$\langle {\bm Q}|{\rm n}_0^a \rangle$
states in Eqs.~(\ref{Man0a}),~(\ref{MaKQQ}) and~(\ref{ExPsiQ}) one can construct
an approximate operator~$\chP_0$ that does not alter odd (or even) states of
the Hamiltonian and strongly affects the states of the opposite symmetry.
Second, the use of simpler kernel in Eqs.~(\ref{MaK1QQ}) and~(\ref{ExKer1}) leads
to qualitatively similar results to those obtained for
the kernel in Eqs.~(\ref{MaKer}) and~(\ref{ExKer}).
Third, the parameter~$\lambda$ can be used as a tool for distinguishing states
having proper or improper symmetry with respect to exchange of all pair of electrons.
Finally, if one uses an approximate kernel in Eqs.~(\ref{MaK1QQ}) or~(\ref{ExKer1}),
then they work correctly for some states only, in above example only for those
with~$|m| \leq 2$.

\section{Discussion \label{Sec_Discussion}}

In this work we introduced the Pauli Exclusion Operator that ensures
appropriate symmetry of multi-electron eigenstate, see
Eqs.~(\ref{ThPs}) and~(\ref{ThPa}).
For two-electron systems we showed three alternative
representations of PEO. In Eqs.~(\ref{ThPe}) and~(\ref{ThPo}) we
expressed PEO in terms of infinite sums over subsets of states
belonging to the spectrum of the Hamiltonian. Using this method we
calculated PEO for isotropic and anisotropic Hooke's atom.

For~$2D$ two-electron systems it is possible to express~$\hP$ in a
closed form in terms of momentum operators, see Eqs.~(\ref{ThPeA})--(\ref{ThP}).
In the position representation~$\hP$ is a nonlocal operator,
and the states of the two-electron Hamiltonian should be calculated
from the nonlocal Yamaguchi equation rather than the Schrodinger
equation, see Eqs.~(\ref{ThPR}) and~(\ref{ThPsiQ}).

In two-electron systems the spectrum of the Hamiltonian contains only
symmetric or antisymmetric states. This is not valid in multi-electron
cases, since for the latter the
solutions of the Schrodinger equation may be symmetric for the exchange of some
pairs of electrons and antisymmetric for the others.
Only application of the Pauli exclusion principle selects states of~$\hH$
that are antisymmetric for exchange of all pairs of electrons.

The PEO can be generalized for multi-electron systems and
it can be defined in two alternative forms: either in terms
of operators~$\hchi_{ij}$
[(see Eq.~(\ref{Mechi})] or by spectral resolution, see Eq.~(\ref{MechP}).
The~$\hchi_{ij}$ operators can be represented as a product of
an infinite power series of position and momentum operators and
electron spins. In this representation PEO depends on
the product of~$\hchi_{ij}$ for all pairs of electrons.
In the second representation~$\chP$ is an operator that includes
all states and energies of the Hamiltonian except
states being antisymmetric with respect to exchange of all pairs of electrons.
For two-electron systems both forms of PEO reduce to results in
Section~\ref{Sec_Theory}. Note that PEO can not be represented in a
closed form for more than two electrons.

Several approximate formulas for~$\chP$ were proposed in Section~\ref{Sec_MultiE}.
The most promising ones for multi-electron systems are based on the approximate
forms of~$\chP_0$ calculated for simpler systems as, e.g., for set of free electrons
in harmonic potential, see Eq.~(\ref{Man0a}).
Another possibility is to treat~$\chP_0$ as a kernel operator that
ensures antisymmetry of the calculated wave function,
see Eq.~(\ref{MaKer}). This kernel may be treated as a mathematical
object without clear physical meaning. Calculated energies and states
of~$2D$ Hooke's atom confirm the effectiveness of these approximations.

It is interesting to compare results obtained with the use of PEO to
variational methods for trial functions taken in form of Slater determinants.
As shown in Eq.~(\ref{MaPEq0}), once the wave function~$|\Psi^a\rangle$
is already anti-symmetrized there is~$\chP|\Psi^a\rangle=0$, and
it is not necessary to introduce PEO.
Variational calculations with the use of trial function in the Slater
form are the most common
method of calculating the energies and states of multi-electron systems. In
practice this method is the best compared to other approaches. The
conclusion is, that for variational calculations with the Slater determinants PEO
is not needed.

However, if one goes beyond variational calculations or if a trial variational
function is not antisymmetric in all pairs of electrons,
then one encounters problem of
ensuring antisymmetry of multi-electron function. This problem could be solved
either ex-post, by eliminating spurious solutions that are not antisymmetric
with respect to exchange of
all pairs or electrons, or by adding PEO to the Hamiltonian
that ensures antisymmetry of resulting wave function. As pointed above,
it seems to be impossible to find exact PEO for arbitrary systems,
but application of approximate forms of POE proposed in Section~\ref{Sec_MultiE}
may be sufficient to obtain a wave function fulfilling
antisymmetry requirement.

The fundamental difference between PEO method and commonly used methods, as e.g.
the configuration interaction (CI) method is as follows. In PEO approach one does not take
any assumption of the wave function but the PEO ensures proper antisymmetry of the
resulting wave function. In the CI method one does not introduce any additional operator, but
assumes the multi-electron wave function as a combination of Slater determinants.
Therefore the PEO method is in some sense 'opposite' to commonly used methods based on
Slater determinants. If both approaches, if one takes exact PEO or exact antisymmetric
trial function the one obtains identical results. However, since in practice one
always uses approximate methods, as e.g. those in Section~\ref{Sec_MultiE}, it may turn
out that in some problems one method is superior to the other. As an example, for~$2D$
Hooke's atom the use PEO gives exact energies and states, see Eq.~(\ref{Exgo}),
but variational method based on
Slater determinants leads to approximate results.

In this work we concentrate on the analysis of the Pauli Exchange Operator for~$2D$
Hooke's atom, which is simpler than Hooke's atom in~$3D$. In the latter
case the Hamiltonian also separates into parts depending on the center-of-mass motion
and the relative motion. The states of the Hooke's atom Hamiltonian in~$3D$ have the
form~$\Psi({\bm r}) = g_{l,n}(r)Y_{l,m}(\Theta,\phi)$, where~$Y_{l,m}(\Theta,\phi)$ are the
spherical harmonics in the standard notation. Functions~$g_{l,m}(r)$
are the solutions of the equation
\begin{equation}
 \label{DiHH}
 \left( -\frac{d^2}{dr^2} - \frac{2}{r}\frac{d}{dr} +\frac{l(l+1)}{r^2} + \frac{1}{r}
 + \frac{k}{4}r^2 \right) g_{l,n}(r) = E_{l,n} g_{l,n}(r),
\end{equation}
where~$l=0,1\ldots$ is angular momentum number and~$E_{l,n}$ are the energies.
Functions~$g_{l,n}(r)$ in Eq.~(\ref{DiHH}) are similar to~$g_{m,n}(r)$ in Eq.~(\ref{ExHg}),
see Figure~\ref{Fig2}. In~$3D$ the transformation~${\bm r} \rightarrow -{\bm r}$ does not
change~$r=|{\bm r}|$ coordinate, but changes the angular functions
\begin{equation} \label{DiY}
 Y_{l,m}(\Theta,\phi) \rightarrow Y_{l,m}(\pi-\Theta,\phi + \pi) = (-1)^lY_{l,m}(\Theta,\phi).
\end{equation}
Then, similarly to~$2D$ case, the states with even~$l$ are symmetric with
respect to exchange of electrons, while those with odd~$l$ are asymmetric.
In~$3D$ one may {\it not} express~$\hP$ in terms of differential operator,
because the transformation~$\Theta \rightarrow \pi-\Theta$ can not be
expressed in terms of translation operator, see Eqs.~(\ref{ThT}) and~(\ref{ThP}).
However, representation of PEO in Eqs.~(\ref{ThPeR})--(\ref{ThPR}) is valid also
in~$3D$ Hooke's atom model.

There exist two systems having two interacting electrons, i.e. the helium atom
and the lithium ion. In these systems the external potential acting on the electrons
is the Coulomb potential of the nucleus. The Schrodinger equations of both systems
do not separate into the center-of-mass and relative motions,
and in order to find eigenvalues or the eigenstates one has to use approximate
methods, e.g. variational calculations, molecular orbital approximations
or perturbation methods~\cite{BetheBook,Tanner2000}.
These methods work correctly
for low energy states but their accuracy decreases for high-energies.
For this reason it is practically impossible
to calculate PEO for helium atom and lithium ion by summating the eigenstates
in Eqs.~(\ref{ThPe}) and~(\ref{ThPo}). However, the results in Figures~1 and~3 suggest
that for both systems the position representation of PEO is also given in
Eqs.~(\ref{ThPeR})--(\ref{ThPR}). Finally, for hypothetical~$2D$ helium atom
PEO is also given by Eq.~(\ref{ThP}).

Let us briefly discuss some issues related to spin part of wave function
for multi-electron systems.
Consider first the three-electron case as e.g. the lithium atom and assume
that the Hamiltonian of the system does not depend on electron spins. In such a case
the wave function of the system is a product of position-dependent and
spin-dependent functions. For three spins there is~$2^3$ spins-combinations, and
they form four quartets and four doublets~\cite{Buchachenko2002}.
The quartet states are symmetric
with respect to exchange of three pairs of spins, but doublets are not, so to ensure
proper symmetry of three-electron wave function a combination of doublets
should be taken. For~$k$-electron system there is~$2^k$ spins combinations, and
for large~$k$ it is practically impossible to treat spins exactly, so one may
either treat them classically, or apply further approximations.

In Section~\ref{Sec_Theory} we assumed spin-independent two-body Hamiltonian.
In real systems one often meets spin-dependent interactions, usually related to
the spin-orbit (SO) coupling. In practical realizations of Hooke's-like
systems in quantum dots the~SO is common, see~\cite{Darnhofer1993,Jacak1997,Poszwa2020}.
In the standard notation
there is~$\hH_{SO} = \alpha \hat{\bm L}\cdot \hat{\bm S}$, and for~$L > 0$.
Then, for~$L > 0$ the wave functions of electrons
do not separate in position-only and spin-only parts and we may not use
the approach in Section~\ref{Sec_Theory}. The general formalism
in Section~\ref{Sec_MultiE} as well as the approximate methods
are valid also for systems with spin-dependent
interactions including~SO.

For two-electron systems in Section~\ref{Sec_Theory} the PEO is defined as an operator
that removes even or odd states from the Hamiltonian spectrum.
Then the function~$|\Psi\rangle$,
being the solution of~$\left(\hH-\hP \right) |\Psi\rangle = E|\Psi\rangle$, includes
odd or even states only.
For multi-electron systems in Section~\ref{Sec_MultiE} the PEO is defined as an operator
that removes antisymmetric states with respect to exchange of all pairs of electrons
from Hamiltonian spectrum. Then the function~$|\Psi\rangle$,
being the solution of~$\left(\hH-\chP \right) |\Psi\rangle = E|\Psi\rangle$, includes
antisymmetric states only. The difference between both definitions is that even or
odd states of two-electron system relate to relative motion of electrons, while
for multi-electron systems the antisymmetry relates to exchange of two electrons
including their positions and spins.

PEO in literature appear previously in calculations of nuclear
matter properties~\cite{Cheon1989,Schiller1999,Suzuki2000}.
In the approach of Ref.~\cite{Suzuki2000} the~$\hat{G}$ matrix
satisfies the Bethe-Goldstone equation
\begin{equation}
 \hat{G} = v + v \frac{\hat{Q}}{\epsilon} \hat{G},
\end{equation}
where~$\hat{G}$ is the reaction matrix,~$v$ is the two-nucleon
interaction,~$\epsilon$ is re-scaled energy and~$\hat{Q}$ is
PEO in nuclear matter which prevents two particles
from scattering into intermediate states with momenta below the Fermi energy.
In some aspects this approach is similar to ours since the authors introduce an
operator responsible for the Pauli exclusion principle, but PEO
in the previous approach excludes some states from real or
virtual scattering. In our approach PEO ensures proper symmetry of
multi-electron wave function.

\section{Summary \label{Sec_Summary}}

In this work we introduce the Pauli Exclusion Operator which ensures
proper symmetry of the states of multi-electron systems with respect to exchange
of each pair of electrons. Once PEO is added to the Hamiltonian,
no additional constraints due to the Pauli exclusion principle
need to be imposed to multi-electron wave function.
PEO is analyzed for two-electron Hamiltonian and we
found its three representations. We concentrated on PEO
in~$2D$ in which it can be expressed in closed form. Some properties of PEO
in~$3D$ and~$1D$ for two-electron states are discussed. PEO are calculated
analytically or numerically for symmetric and antisymmetric Hooke's atoms.
We generalized PEO for multi-electron systems; its two alternative forms
are obtained. Several approximations of PEO to multi-electron systems were
derived. Kernel-based methods were proposed, and they seem to be most promising
approximations of PEO for practical calculations. It is shown that once the
wave function~$|\Psi^a\rangle$ is already antisymmetric with respect to exchange of
all pairs of electrons,~$\chP|\Psi^a\rangle$ identically vanishes. For
this reason, in variational calculations employing trial functions in the form of
Slater determinants there is no need to introduce PEO. However, if one goes beyond
variational calculations, one should introduce PEO to ensure antisymmetry of
the resulting wave functions. We believe that the approach based on exact,
approximate or kernel forms of~PEO may be useful in calculating
energies and states of multi-electron systems.

\appendix

\section{Auxiliary identities} \label{AppA}

The spin-exchange operator~$\hSig_{ij}$ is defined
by its action on four two-spin states
\begin{eqnarray}
 \hSig_{ij} | \uparrow   \uparrow   \rangle = | \uparrow   \uparrow   \rangle,\\
 \hSig_{ij} | \uparrow   \downarrow \rangle = | \downarrow \uparrow   \rangle, \\
 \hSig_{ij} | \downarrow \uparrow   \rangle = | \uparrow   \downarrow \rangle, \\
 \hSig_{ij} | \downarrow \downarrow \rangle = | \downarrow \downarrow \rangle.
\end{eqnarray}
Operator~$\hSig_{ij}$ in Eq.~(\ref{MeSig}) satisfies all above equations.

In~$3D$ the particle exchange operator is given in Eq.~(23) of Ref.~\cite{Schmid1979}
and for completeness we quote this expression
\begin{eqnarray} \label{AppA_hPi3}
 \hPi_{ij} = \sum_{n=0}^{\infty}&& \left( \frac{1}{n!}\right) \left(\frac{i}{\hbar}\right)^n
 \sum_{l=0}^n \sum_{m=0}^l \left(\begin{array}{c} n \\ l \end{array}\right)
 \left(\begin{array}{c} l \\ m \end{array}\right)
 \nonumber \\
 &\times & \left(\hat{p}_{jx}-\hat{p}_{ix} \right)^{n-l}\left( \hat{r}_{jx}-\hat{r}_{ix}\right)^{n-l}
 \nonumber \\
 &\times & \left(\hat{p}_{jy}-\hat{p}_{iy} \right)^{l-m} \left( \hat{r}_{jy}-\hat{r}_{iy}\right)^{l-m}
 \nonumber \\
 &\times & \left(\hat{p}_{jz}-\hat{p}_{iz} \right)^m \left( \hat{r}_{jz}-\hat{r}_{iz}\right)^m.
 \end{eqnarray}
By taking limit~$\Delta \hat{x}_{ij} \rightarrow 0$ in Eq.~(\ref{AppA_hPi3}) one obtains
the particle exchange operator in~$2D$
\begin{eqnarray} \label{AppA_hPi2}
 \hPi_{ij} = \sum_{n=0}^{\infty}&& \left( \frac{1}{n!}\right) \left(\frac{i}{\hbar}\right)^n
 \sum_{l=0}^n \left(\begin{array}{c} n \\ l \end{array}\right)
 \nonumber \\
 &\times & \left(\hat{p}_{jx}-\hat{p}_{ix} \right)^{n-l}\left( \hat{r}_{jx}-\hat{r}_{ix}\right)^{n-l}
 \nonumber \\
 &\times & \left(\hat{p}_{jy}-\hat{p}_{iy} \right)^{l} \left( \hat{r}_{jy}-\hat{r}_{iy}\right)^{l}.
 \end{eqnarray}
Alternative expressions for~$\hPi_{ij}$ is given in~\cite{Grau1981}.

\section{Shooting method \label{AppB}}

The eigenenergies and eigenstates of Hooke's atom in Eq.~(\ref{ExHg})
are found using shooting method~\cite{NRecBook}.
As the initial guesses for the energies we use those
of two-dimensional harmonic oscillator
equal to~$E_n= \sqrt{(k/4)}(2n+1)$,~$n=0,1,..\ldots$m and~$k=4$.
Then we iteratively bracket the true energies of~$\hH$ by analyzing
behavior of~$g_{m,n}(r)$ at large~$r$. The advantage of the shooting method is
that it is equally accurate for low and high energy states.
Only functions with~$m\ge 0$ were calculated since~$g_{-m,n}(r) = g_{m,n}(r)$.
We tabulate normalized states of Eq.~(\ref{ExHg}) from~$n=1$ (ground state)
to~$n=250$ and from~$m=0$ to~$m=16$.

We solve Eq.~(\ref{ExHg}) using DVERK procedure
which is~$6-$th order Runge-Kutta method~\cite{Verner1971,HairerBook}.
The accuracy of calculations has been verified by checking the orthogonality
of all pairs of~$g_{m,n}(r)$ and~$g_{m',n'}(r)$ functions
with~$m=m'$ and~$n\neq n'$. In each case the accuracy below~$10^{-5}$ has
been obtained.

For small~$r$ there is:~$g_{0,n}(r) \simeq c_0(1+r)$ with~$c_0>0$,
and the initial conditions for DVERK procedure are:~$g_{0,n}(0)=1$,~$g_{0,n}'(0)=h$,
where~$h$ is the integration step, and~$h \simeq 0.001 - 0.01$~r$_B$.
For~$m>0$ and small~$r$ there is:~$g_{m,n}(r) \propto r^{m}$, and
the initial conditions for DVERK procedure are:~$g_{m,n}(0)=0$,~$g_{m,n}'(0)=mh^{m-1}$.
For large~$m$ the last condition
is unstable numerically, and it is replaced
by:~$g_{m,n}(r_0)=g_c$,~$g_{m,n}'(r_0)= m g_c/r_0$,~$g_c \simeq 10^{-5}$,
and~$g_{m,n}(r)=0$ for~$r < r_0$.
Here~$r_0 >0$ and its values for~$g_{m,n}(r)$
are obtained by analysis of~$g_{m-1,n}(r)$ for small~$r$.
Generally,~$r_0$ gradually increases with~$m$.

\end{document}